\documentclass[twocolumn,aps,groupedaddress,superscriptaddress,showpacs,pra,floatfix]{revtex4}

\usepackage{graphicx}
\usepackage{dcolumn}
\usepackage{bm}
\usepackage{epsfig}
\usepackage{dsfont,psfrag}
\usepackage{color}
\usepackage{amsfonts}
\usepackage{amsmath}
\usepackage{amssymb}
\usepackage{nicefrac}

\newcommand{\LF}{L_{\text F}}
\newcommand{\kF}{k_{\text F}}

\newcommand{\dd}{\, \mathrm d}

\newcommand{\pdag}{{\phantom\dagger}}
\newcommand{\hc}{\text{h.c.}}
\newcommand{\be}{\begin{eqnarray}}
\newcommand{\ee}{\end{eqnarray}}

\begin{document}

\title{Density profile of interacting Fermions in a one-dimensional optical trap}
\author{Stefan A.~S\"{o}ffing}
\affiliation{Dept.~of Physics and Research Center OPTIMAS and Graduate School MAINZ/MATCOR, Univ.~Kaiserslautern, D-67663 Kaiserslautern, Germany }
\author{Michael Bortz}
\affiliation{Fraunhofer ITWM, 67663 Kaiserslautern, Germany}
\author{Sebastian Eggert}
\affiliation{Dept.~of Physics and Research Center OPTIMAS and Graduate School MAINZ/MATCOR, Univ.~Kaiserslautern, D-67663 Kaiserslautern, Germany }
\date{\today}

\begin{abstract}
The density distribution of 
the one-dimensional Hubbard model in a harmonic trapping potential 
is investigated in 
order to study the effect of the confining trap. 
Strong superimposed oscillations are always present on top of a uniform density 
cloud, which show universal scaling behavior as a function of increasing interactions.
An analytical formula is proposed on the basis of bosonization, which describes
the density oscillations for all interaction strengths.
The wavelength of the dominant oscillation changes with interaction, which 
indicates the crossover to a spin-incoherent regime.  
Using the Bethe ansatz 
the shape of
the uniform fermion cloud is analyzed in detail, 
which can be
described by a universal scaling form.

\end{abstract}


\maketitle

Ultra-cold gases in optical traps and lattices have become a 
promising tool for simulating strongly correlated systems
with a full control of all relevant parameters \cite{review}.
While the first simulations were mostly made on bosonic setups, ultra-cold 
fermions are by now also well established \cite{fermions}.  In order to simulate
interacting electron systems such as Hubbard-type models, fermionic atoms with
two different hyperfine states are used in order to represent
the two spin channels \cite{fermions}.
It is therefore possible to test theoretical predictions even for 
systems that are less common
or hard to produce in nature, such as perfectly clean isolated one-dimensional (1D)
quantum wires.
However, the experimental setup will always possess a smoothly varying potential due to the
intensity profile of the laser beams, usually forming a harmonic confinement.

Recent experimental developments have made it possible to locally 
probe the density profile of ultra-cold atomic condensates
directly in space using optical microscopy \cite{kuhr} or electron beam
scanning \cite{ott}.
In this work we therefore want to provide a detailed theoretical  
quantitative analysis of the 1D fermion density 
profile as a function of the interaction strength and the confining potential, which in 
turn can be used to analyze interaction effects from the experimental signals.

The fermion density can generally 
be characterized in terms of two distinct features, namely the 
overall size of the cloud on the one hand and superimposed density oscillations on the 
other hand.
From works on 
quantum wires and quantum dots it is well known that 
density oscillations may appear from reflections at 
sharp edges and boundaries, which are
due to interference (Friedel oscillations) and/or localization (Wigner
crystallization) \cite{friedel,mueller,Soffing09,Bedurftig98,rommer,White02}.
However, it is not \textit{a priori} clear how these oscillations 
are modified if a harmonic potential is present as a confinement.
In this work we now show that the oscillations remain strong in a harmonic trap
with interactions, despite the lack of any sharp edges which may 
cause Friedel oscillations.  
In a pioneering work \cite{schulz} from 1993, 
Schulz predicted $4k_F$ density correlations to dominate, which he called a one-dimensional 
''Wigner crystal'', that only occurs
in case that the interaction parameter takes on rather 
extreme values, which cannot be reached in a short-ranged Hubbard model even for 
infinite $U$.  In contrast to this expectation, we now find a surprising 
crossover towards rather strong $4k_F$ 
''Wigner oscillations'' in a trap even for intermediate short-range interactions.
Moreover, both Friedel and Wigner oscillations can actually
be very well analyzed with the help of an analytic formula on the basis of a bosonization 
approach.  The overall size and shape of the density cloud is also analyzed in quantitative 
detail, which follows a universal scaling form. 

We consider the standard 1D Hubbard Hamiltonian with an external trapping potential 
\begin{eqnarray}
	H  =   \sum_{x} \sum_{\sigma=\uparrow,\downarrow}
& &  \!\!\!\!\!\!\!\!	\left( -J(\psi_{\sigma, x}^\dagger \psi_{\sigma, x+1}^\pdag + \hc ) \right.
\nonumber \\
 & &\left. +  (\mu_0 +\omega^2 x^2) \, n_{\sigma, x}
		+ \frac{U}{2} n_{\sigma, x} n_{\sigma, x}
\right) 
	\label{H}
\end{eqnarray}
in the limit of 
large particle separations (small densities)
relative to the lattice spacing. In this limit \cite{muth} the Hamiltonian can also be approximated
by the continuous problem of fermions 
with contact interactions
\begin{equation}
H= \sum_{n=1}^N \left( -J \ \partial_{x_n}^2 + \omega^2 x_n^2 + 
\sum_{m=1}^N \frac{U}{2} \delta(x_n-x_m)\right), \label{Hcont}
\end{equation}
where we assume a non-magnetic state with fixed particle number 
$N =2N_\uparrow=2N_\downarrow$.  The lattice spacing and the hopping $J$
are the natural units for this problem which are set to unity in what follows.
The condition for large particle separation corresponds to ${N\omega} \ll 1$.
It should be noted that the opposite limit of small particle separations
(i.e.~of the order of the lattice spacing) has been studied
elsewhere and is governed by a transition to a 
Mott-insulator \cite{Rigol03, Rigol04,Xianlong06-1, Xianlong08, Campo05,perturb,chen}.
A good qualitative understanding of the continuous problem  has been achieved 
using density functional methods \cite{1d,kecke}.

In order to simulate the Hamiltonian in Eq.~(\ref{H}) we use the 
numerical
density matrix renormalization group (DMRG) \cite{white}.
While the DMRG is 
best suited for homogeneous systems with open boundary conditions, it is 
also possible to implement the algorithm to describe inhomogeneous traps 
as long as the
actual system size in the simulation is much larger than 
the spread of the confined
fermions.

\begin{figure}
	\includegraphics[width=0.5\textwidth]{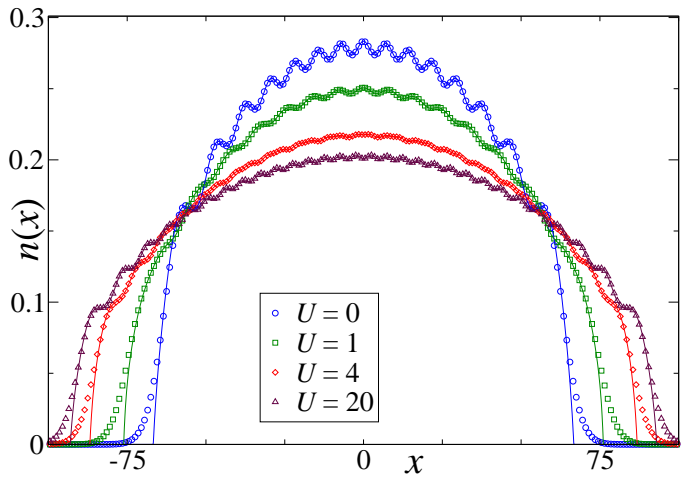}
	\caption{(Color online) DMRG data for the fermion density $n(x)$ 
in a harmonic trap with $N= 30$ and $\omega^2=4 \cdot 10^{-5}$
(points) compared to the analytical 
approximation in  Eq.~\eqref{den_fitformula} (solid lines).}
	\label{fig:den}
\end{figure}

In Fig.~\ref{fig:den}
typical density distributions from DMRG are plotted
for different $U$ in a trap with $N=30$ particles and  $\omega^2=4 \cdot 10^{-5}$, 
showing a localized
fermion cloud with superimposed characteristic oscillations of different wavelengths.  
An analytical approximation
to the data is also shown (solid lines), which will be derived in the following
using bosonization and Bethe ansatz methods.

In order to understand the behavior of the density let us first consider the
non-interacting case.  In the continuous limit 
the wave-functions of the single-particle oscillator levels are given by 
	$h_n(x) = \sqrt{\tfrac{1}{2^{n} n!}} \left(\tfrac{\omega}{\pi}\right)^{1/4} \, e^{-\omega x^2 /2} \, H_n(\sqrt{\omega} x)$,   
where $H_n(x)$ denotes the $n$-th Hermite polynomial.
The ground state density distribution at $U=0$ can be calculated as the sum
over the filled Fermi sea of oscillator levels
\begin{equation}
	n(x) = 2 \sum_{n=0}^{N/2-1} |h_n(x)|^2
	\label{den_nonint}
\end{equation}
for a system containing $N$ electrons.
Using an expansion around the center of the trap, this function can be 
well described by a simple closed formula \cite{Gleisberg00,Butts97,Silvera81}
\begin{equation}
    n(x)
    \approx n_0(x)
    - \frac{(-1)^{N/2}}{\pi\LF} \frac{\cos\left(2\kF(x) x \right)}{1-x^2/\LF^2 }
    \label{den_approx}
    ,
\end{equation}
for $|x|\leq L_F$, where  the density cloud is given by 
\begin{equation}
	n_0(x) = \frac{ 2\omega L_F}{\pi} \sqrt{1-x^2/\LF^2},
	\label{den_semiclas}
\end{equation}
with a Thomas-Fermi size of $L_F=\sqrt{N/\omega}$ for $U=0$.
The result in Eq.~(\ref{den_approx}) resembles the corresponding expression 
for Friedel oscillations in a 1D box \cite{friedel}, which also decay proportional to the
reciprocal distance from the turning points $\pm L_F$.
The slowly varying part of the density $n_0(x)$, Eq.~\eqref{den_semiclas}, 
replaces the normally constant filling $n_0$.
The period of the oscillations is related to the filling, so that the
wavevector also becomes position dependent and is given 
by the \emph{non-local} expression
\begin{eqnarray}
	2\kF(x) x & = &  \pi \int_0^x n_0(y) \,dy \\ 
	 & = &   \omega L_F \left[ x \sqrt{1-x^2/\LF^2} + L_F \arcsin(x/\LF) \right],
\nonumber 	\label{kf}
\end{eqnarray}
which follows from an expansion of the summed up oscillator wave-functions 
in Eq.~(\ref{den_nonint}).  Note that the integration in Eq.~(\ref{kf}) is 
similar 
to the usual WKB approximation, where the local momentum $k(x) \approx \pi n(x)/2$
is integrated in space in order to predict the behavior in a changing potential.
For constant filling the expression in Eq.~(\ref{kf})
reduces to the usual relation  $\kF = \frac{\pi}{2} n_0$. 

\begin{figure}
	\includegraphics[width=0.5\textwidth]{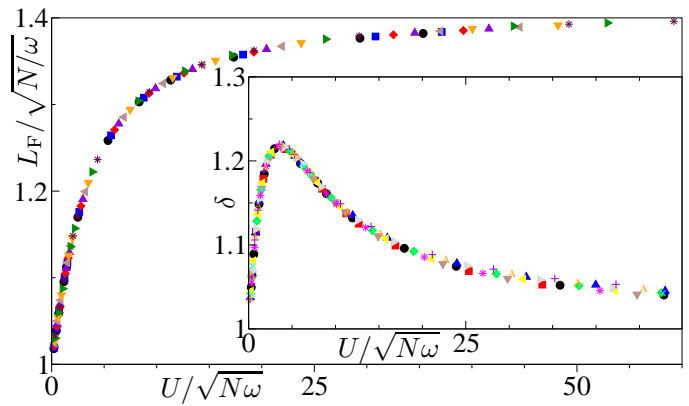}
	\caption{(Color online) The width $\LF(U)$ of the fermion cloud in 
the trap as determined from the 
local Bethe ansatz density as a function of the scaling variable
$U/\sqrt{N\omega}$.  The different symbols (colors) correspond to many different choices 
 of $U \in [0,20]$,    $\omega^2 \in [10^{-5}, 16\times 10^{-5}]$ and  $N \in [10,70]$. 
Inset: Effective exponent $\delta$ 
determined from a fit of $n_0^\delta(x)$ to the local Bethe ansatz density. }
	\label{fig:LF}
\end{figure}
Before we analyze the oscillations in the presence of interactions, 
let us first consider the overall shape of the 
density cloud $n_0(x)$ for repulsive interactions $U>0$.
For a translationally invariant 1D Hubbard model, the density is known
as a function of $U$ and chemical potential $\mu$ from the
non-trivial solution of the Bethe ansatz equations \cite{book}.
This is in sharp contrast to bosonic systems where the density is well described by 
a mean field  approach, which can even 
be applied locally to non-uniform systems with the help of the Gross-Pitaevskii equation.
One promising approach for the non-homogeneous fermion system may be to use the exact 
solution. In particular,  
if the external potential is slowly varying, the Bethe ansatz
density for the chemical potential $\mu(x)=\mu_0+\omega^2 x^2$  could 
be a good approximation for each location $x$ in the trap. 
Since strong long-range correlations exist, it is not {\it a priori} clear
if such a local Bethe ansatz 
approximation with a translationally invariant system at each point 
is appropriate, but it agrees very well with our DMRG data for all $U$.
This is especially surprising near the edges where the filling is low and also the 
discrete energy spectrum should play a role.
At $U=0$ this approach corresponds to the density in Eq.~(\ref{den_semiclas}).
At $U\to \infty$ the interaction induces a  Pauli principle between
spin-up and down electrons, so that the system contains effectively twice
$N=N_\uparrow + N_\downarrow$ non-interacting spin-incoherent particles, which results again
in the density in Eq.~(\ref{den_semiclas}), but with $L_F$ multiplied by a factor of
$\sqrt{2}$.  For intermediate $U$ the total size of the cloud therefore becomes
interaction dependent with $\sqrt{N/\omega} \leq L_F(U) \leq \sqrt{2 N/\omega}$.
We find that the
interaction dependence of the effective 
size $L_F(U)$ in fact follows a universal scaling behavior as 
a function of $U/\sqrt{N\omega}$ as shown in Fig.~\ref{fig:LF},
which is related to the scaling behavior of the Bethe ansatz equations with $U/n_0$
at low filling $n_0$ \cite{book,Soffing09}.
The Bethe ansatz density at intermediate $U$ does not follow exactly the simple expression in
Eq.~(\ref{den_semiclas}), but can be approximated by the normalized shape if it is 
raised by a small exponent  of $\delta \alt 1.3$ as shown in the inset in Fig.~\ref{fig:LF}.
It should be noted that Fig.~\ref{fig:LF} gives a quantitative estimate of the 
screening cloud for all interaction strengths $U$ and trap parameters as long as 
$N\omega \ll 1$.

\begin{figure}
	\includegraphics[width=0.5\textwidth]{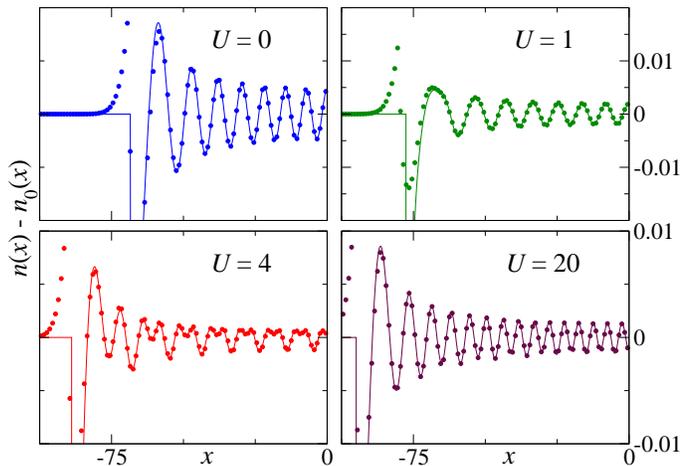}
	\caption{(Color online) Density oscillations around the slowly varying part $n_0(x)$
for $N=30$ and $\omega^2=4 \cdot 10^{-5}$ compared to the analytical form in Eq.~(\ref{den_fitformula}).} 
	\label{fig:osc}
\end{figure}
We can now subtract the 
Bethe ansatz estimate for the 
slowly-varying part of the density $n_0(x)$ from the DMRG data 
in order to analyze the oscillations as shown in Fig.~\ref{fig:osc}.
For weak interactions the Friedel-type oscillations in Eq.~(\ref{den_approx}) 
can clearly be seen.
At intermediate interactions $U=4$ {\it two} dominant wavevectors can be 
observed.  At larger $U$ the faster oscillations dominate, corresponding to exactly
one density maximum per particle, which is one typical signature of 
Wigner crystal oscillations.
The oscillating signal 
is quite sensitive to the 
estimate of the uniform density, which has been subtracted.  
However, as can be seen in Fig.~\ref{fig:osc} the oscillations are 
symmetric in the entire trap without any visible bias towards positive or 
negative values, which shows that the local Bethe ansatz estimate 
works well.

A natural tool for calculating the density oscillations with the help of 
correlation functions in one-dimensional systems
is bosonization \cite{Soffing09}.  In the presence of a trapping potential, 
bosonization has been considered for interacting spinless 
fermions before \cite{Wonneberger01,Xianlong03, Artemenko04, Xianlong04}.
For the spinful case we 
will now derive
the central definitions of the bosonic creation and annihilation operators.
Instead of the usual left- and right-moving  bosons, 
only one bosonic field each for spin and charge ($\nu=\mathrm{ c,s}$) is defined 
\be
\phi_\nu(u) = \phi_\nu^0 + \frac{\hat{N}_\uparrow \pm \hat{N}_\downarrow}{\sqrt{2}} u + 
\sum_{n=1}^\infty\left( \tfrac{i}{\sqrt{n}} e^{-inu} b_{\nu,n}^\pdag  + \hc\right)
\label{bosfield}
\ee
where the bosonic annihilation and creation operators 
\begin{equation}
b^\dagger_{\nu,n} = \frac{1}{\sqrt{2n}}\sum_m (c_{\uparrow,m+n}^\dagger 
c_{\uparrow,m}^{\phantom{\dagger}} \pm 
c_{\downarrow,m+n}^\dagger c_{\downarrow,m}^{\phantom{\dagger}})
\end{equation}
are expressed in terms of fermion operators 
$c_{\sigma,m}^\dagger$ of  the $m$-th oscillator mode
 that are extended 
to include non-physical states $m<0$ (with $\pm$ corresponding to $\nu= \mathrm{ c,s}$ respectively). 
The number operators $\hat{N}_\uparrow \pm \hat{N}_\downarrow$ are canonical conjugate to 
the zero modes $\phi_{\nu}^0$.  
The auxiliary 
variable $u\in  [-\pi,\pi[$ should not be confused with the position $x$.
Following the usual steps of the bosonization 
procedure \cite{egg07}, 
it is then easy to show that free-particle excitations relative to the Fermi edge 
are reproduced by the bosonic Hamiltonian
\begin{equation}
H = \hbar \omega \left(\sum_{\nu,n>0} n b_{\nu,n}^\dagger b_{\nu,n}^{\phantom{\dagger}} 
+ (\hat{N}_\uparrow^2+\hat{N}_\downarrow^2)/2\right)
\end{equation}
The Fourier transformation of the oscillator levels defines a canonical 
auxiliary fermion field which can be bosonized as a vertex operator as usual
\be
\tilde\psi^\dagger_\sigma(u) \equiv \frac{1}{\sqrt{2 \pi}} 
\sum_{m} e^{i mu} \,c_{\sigma,m}^\dagger
\propto  e^{-i (\phi_c(u)\pm \phi_s(u))/\sqrt{2} }.
	\label{psia}
\ee
The auxiliary field 
 can in turn be used to give a non-local expression of the physical fermion fields 
in terms of the bosons
\be
	\psi^\dagger_\sigma(x) & \equiv & \sum_{n=0}^\infty h_n(x) \, c_{\sigma,n}^\dagger 
\nonumber \\
 & = &   
   \frac{1}{\sqrt{2\pi}} 
	\int_{-\pi}^{\pi}\, \sum_{n} h_n(x) \, e^{- i n u} \tilde\psi^\dagger_\sigma(u)
{\dd u}
	\label{fermion-creator-bos}
	.
\ee
This expression can be made approximately 
local in $x$ by noticing that the wave-functions near the Fermi edge 
oscillate roughly as a function 
of  $n$ with $h_n(x) \sim \cos (n \arccos(x/L_F))$.  Therefore, 
$\psi^\dagger_\sigma(x) \sim e^{-i (\phi_c(u)\pm \phi_s(u))/\sqrt{2}}
+ e^{-i (\phi_c(-u)\pm \phi_s(-u))/\sqrt{2}}$ with 
 $u \approx \arccos(x/L_F)$,
which also leads to densities in terms of derivatives of the 
boson field \cite{Wonneberger01,Xianlong03, Artemenko04, Xianlong04}.
Without interactions this bosonization approximation is in fact more accurate than
for translational invariant systems due to the linear oscillator spectrum.
However, spinful interactions become quite complicated in the bosonized language,
since general scattering
terms appear that are not even momentum conserving.

\begin{figure}
	\includegraphics[width=0.5\textwidth]{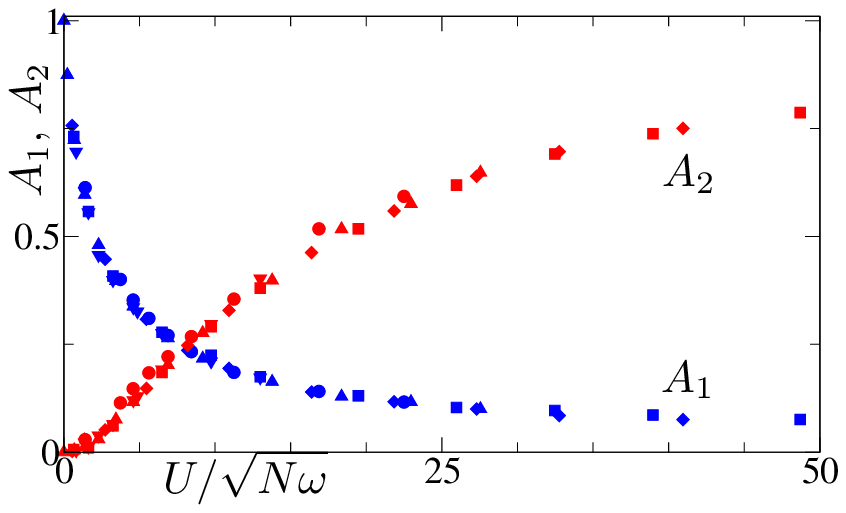}
	\includegraphics[width=0.5\textwidth]{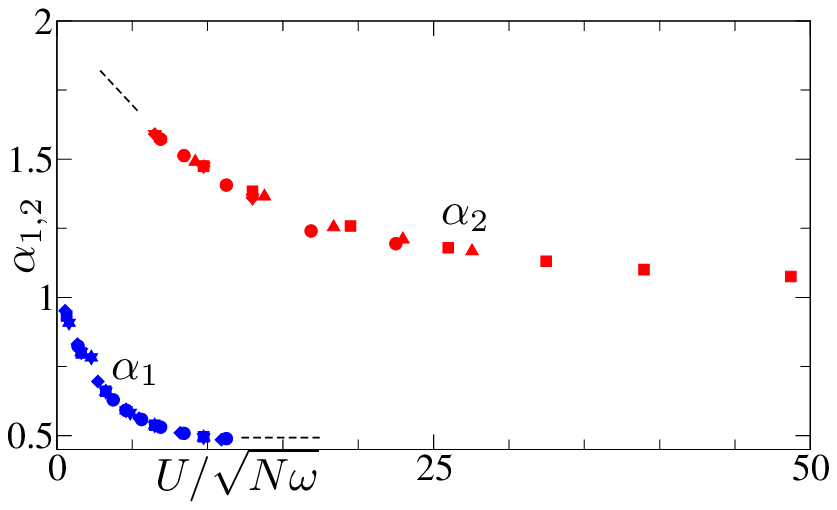}
	\caption{(Color online) Amplitudes $A_{1,2}$ (top) and exponents $\alpha_{1,2}$ (bottom) as determined by fitting
	Eq.~\eqref{den_fitformula} to DMRG data for different choices of $U$, $\omega$, and $N$.}
	\label{fig:fit}
\end{figure}
Although we are not able to solve the system by a simple Bogoliubov transformation, 
the bosonization picture is useful since 
it is reasonable to expect that the
leading instabilities are again a $2k_F$ Friedel and a 
$4k_F$ Wigner oscillation as for the translational 
invariant system \cite{schulz,Soffing09} albeit with a changing 
wavevector along the trap according to Eq.~(\ref{kf}).  
We can therefore generalize Eq.~(\ref{den_approx}) analogously to the Hubbard model with fixed boundary
conditions and propose a general ansatz for the density in the trap
for $|x|\leq L_F$
\begin{eqnarray}
	n(x)
	= n_0(x)
	& - &  A_1 \frac{(-1)^{N/2}}{\pi \LF} \frac{\cos\left( 2\kF(x) \, x \right)}{[1-x^2/L_F^2]^{\alpha_1}} \nonumber \\ 
	 & - &    A_2 \frac{1}{2\pi \LF} \frac{\cos\left( 4\kF(x) \, x \right)}{[1-x^2/L_F^2]^{\alpha_2}},
	\label{den_fitformula}
\end{eqnarray}
where $L_F(U)$ is given in Fig.~\ref{fig:LF} and $\kF(x)$ is given in Eq.~(\ref{kf}) as 
a function of $L_F$.
The amplitudes $A_{1,2}$ and the exponents $\alpha_{1,2}$ are unknown and 
have to be determined from fitting the DMRG data.
As can be seen in Fig.~\ref{fig:osc} this formula fits the data extremely well 
even to within the last oscillation near the edge.
In Fig.~\ref{fig:fit} the results for the amplitudes and the exponents are shown which 
again follow a scaling law as a function of $U/\sqrt{N\omega}$.  
For smaller amplitudes $A_{1,2} \alt 0.2$ the corresponding exponents in Fig.~\ref{fig:fit}
could no longer be 
accurately determined.
A clear increase of the faster Wigner-crystal oscillations $A_2$ can be 
seen with increasing $U$. 
At the same time the slower Friedel-type oscillations $A_1$ are suppressed.
Both decay exponents generally decrease with increasing interactions which 
is qualitatively similar to the translational invariant Hubbard model \cite{book}.

In conclusion we have analyzed the detailed behavior of the 
fermion density of the one-dimensional
Hubbard model in a harmonic trap with the help of bosonization and the Bethe ansatz.  
The proposed analytical formula
in Eq.~(\ref{den_fitformula})
and the scaling behavior of the parameters in Fig.~\ref{fig:LF} and Fig.~\ref{fig:fit}
provide very accurate predictions  
for the position dependent density $n(x)$ as a function of 
arbitrary interaction strengths $U$ and trap parameters in the limit $N\omega \ll 1$.
Significant deviations can only be observed in the last oscillations near the edge of the
density cloud.
The overall density $n_0(x)$ follows a local Bethe ansatz approximation and the
oscillations in the trap remain strong despite the lack of any hard-wall boundary conditions.
A crossover from slower Friedel oscillations to faster Wigner crystal oscillations can 
be observed with
increasing $U$. 
We hope that our results will be useful in the analysis of future experiments on
ultracold fermions in a one-dimensional trap with local resolution. 
At the same time the good fit to the
proposed analytical formula in Eq.~(\ref{den_fitformula}) strongly suggests 
that the problem can be solved by further analyzing the
bosonization formulas in Eqs.~(\ref{bosfield}-\ref{fermion-creator-bos}) in the presence of interactions, 
which might inspire future research 
on the topic.
%
%

We are thankful for useful discussions with I. Schneider and A. Struck.
This work was supported by the DFG via 
the SFB/Transregio 49 and the MAINZ school of excellence.
%
%
%
%



%
%
%
%
%
%
%

\end{document}